\input harvmac.tex
\def\exp{{\rm exp}}

\def\frac#1#2{{#1\over#2}}

\lref\FATEEV{Fateev, V.A.: to be published}

\lref\KouM{Koubek, A. and Mussardo, G.: 
On the operator content of the sinh-Gordon model.
Phys. Lett. {\bf B311}, 193-201 (1993)}

\lref\ZZ{Zamolodchikov, A.B. and Zamolodchikov, Al.B.:
Structure Constants and Conformal Bootstrap in
Liouville Field Theory.
Nucl. Phys. {\bf B477}, 577-605   (1996) }

\lref\Zarn{Zamolodchikov, Al.B.: Mass scale
in the Sine-Gordon model and its
reductions. Int. J. Mod. Phys. {\bf A10}, 1125-1150  (1995) }

\lref\GZ{Ghosal, S. and Zamolodchikov, A.B.:
Boundary S-matrix and boundary state in two-dimensional integrable
quantum field theory.
Int. J.  Mod. Phys. {\bf A9}, 3841-3885 (1994)}

\lref\Warn{Fendley, P., Saleur, H. and Warner, N.P.:
Exact solution of a massless scalar
field with a relevant boundary interaction.
Nucl. Phys. {\bf B430}, 577-596 (1994)} 

\lref\Fen{Fendley, P., Ludwig, A.W.W.  and Saleur, H.:
Exact non-equilibrium transport through point contacts in
quantum wires and fractional quantum Hall devices.
Phys. Rev. {\bf B52},
8934-8950 (1995)}

\lref\BLZ{Bazhanov, V.V., Lukyanov, S.L. and Zamolodchikov, A.B.:
Integrable structure of conformal field theory.
2. Q-operator and DDV equation.
Preprint CLNS 96/1405, LPTENS 96-18,
\#hepth  9604044}

\lref\Kane{Kane, C.L. and Fisher, M.P.A.:
Transmission through barriers and resonant tunneling in
an interacting one-dimensional electron gas.
Phys. Rev.  {\bf B46}, 15233-15262 (1992)} 

\lref\Moon{Moon, K., Yi, H., Kane, C.L., Girvin, S.M. and
Fisher, M.P.A.: Resonant tunneling between quantum Hall
edge states. Phys. Rev. Lett. {\bf 27}, 4381-4384 (1993) }

\lref\SalF{Fendley, P., Lesage, F. and Saleur H.:
A unified framework for the Kondo problem and
for an impurity in a Luttinger liquid.
J. Stat. Phys. {\bf 85}, 211-249 (1996)}

\lref\Schmid{Schmid, A.:
Diffusion and localization in a dissipative
quantum system.
Phys. Rev. Lett. {\bf 51}, 1506-1509 (1983)}

\lref\leggett{Caldeira, A.O.
and Legget, A.J.: Influence
of dissipation on quantum tunneling in
macroscopic systems. Phys. Rev. Lett. 
{\bf 46}, 211-214 (1981)\semi
Caldeira, A.O.
and Legget, A.J.:
Path integral approach to quantum
Brownian motion.
Physica  {\bf A121}, 587-616 (1983)}

\lref\Callan{Callan, C.G. and Thorlacius, L.:
Open string theory as dissipative quantum 
mechanics. Nucl. Phys. {\bf B329}, 117-138 (1990)}

\lref\Fisher{Fisher, M.P.A. and Zwerger, W.:
Quantum Brownian motion in a
periodic potential.  Phys. Rev.
{\bf B32}, 6190-6206 (1985)}

\lref\LZ{Lukyanov, S. and Zamolodchikov, A.:
Exact expectation values of local fields
in quantum sine-Gordon model.
Preprint CLNS 96/1444, RU-96-107,
\#hepth  9611238}

\Title{\vbox{\baselineskip12pt\hbox{CLNS 97/1465}
\hbox{RU-97-04}                
\hbox{hep-th/9702190}}}
{\vbox{\centerline{
Expectation values of boundary fields}
\vskip6pt
\centerline{in the boundary sine-Gordon model}}}
 
\centerline{Vladimir Fateev$^{1,4}$,
Sergei Lukyanov$^{2,4}$,}
\centerline{
Alexander Zamolodchikov$^{3,4}$
and Alexei Zamolodchikov$^{1}$}

\centerline{}
\centerline{$^1$Laboratoire de Physique
Math\'ematique, Universit\'e de Montpellier II}
\centerline{ Pl. E. Bataillon,  34095 Montpellier, FRANCE}
\centerline{$^2$Newman Laboratory, Cornell University}
\centerline{ Ithaca, NY 14853-5001, USA}
\centerline{$^3 $Department of Physics and Astronomy,
Rutgers University}
\centerline{ Piscataway,
NJ 08855-0849, USA}
\centerline{and}
\centerline{$^4$L.D. Landau Institute for Theoretical Physics,}
\centerline{Chernogolovka, 142432, RUSSIA}

\centerline{}
\centerline{}
\centerline{}
 
\centerline{\bf Abstract}

We propose an explicit expression for  vacuum expectation values of
the boundary field $e^{ia\varphi_{B}}$ in the boundary sine-Gordon model
with zero bulk mass. This expression agrees with  known exact 
results for the boundary free energy and with  perturbative
calculations.
 
\Date{February, 97}
\vfill
 
\eject

In this paper we study the
so called boundary sine-Gordon
model  with zero
bulk mass. This is a two-dimensional quantum field theory defined by the 
Euclidean action
\eqn\basa{\eqalign{\cal
A}_{BSG}={1\over 4 \pi}
\int_{-\infty}^{\infty}dx\int_{0}^{\infty}dy\
\Big\{
\big(\partial_{y}\varphi\big)^2 + \big(\partial_{x}\varphi\big)^2\Big\}
-2\mu_{B}
\int_{-\infty}^{\infty}dx \,\cos\big(\beta\varphi(x,0)\big)\ .}
Here $x,y$ are coordinates on the  Euclidean half-plane $y\geq 0$ and 
$\varphi(x,y)$ is a scalar field. Interaction is present only at the
boundary $y=0$ and it is controlled by two parameters $\beta$ and $\mu_B$.
This model can be understood as a conformal field theory -  a
free Bose field
with free boundary condition at $y=0$ - perturbed by a boundary operator
$2\cos(\beta\varphi_{B})(x)$ of the dimension $\Delta = \beta^2$.
Here we use the notation 
$$\varphi_{B}(x)\equiv 
\varphi(x,0)\ .$$
Correspondingly, we assume that this boundary operator 
is normalized according to the following simple asymptotic form of its
two-point correlation function
\eqn\nodt{\langle\,  2\cos(\beta\varphi_{B})
(x)\,2\cos(\beta\varphi_{B})(x')\, 
\rangle \to 2\,
|x-x'|^{-2\beta^2} \qquad {\rm as} \qquad |x-x'| \to 0\ .}
Under this normalization the parameter $\mu_B$ has the dimension
$[\, mass\, ]^{1-\beta^2}$. 

This model attracted much interest recently
in connection with the
impurity problem in the 1D Luttinger model\ \Kane, \Moon. 
Quantum Brownian motion of a
one-dimensional particle in a 
periodic potential is another interesting
application
of this model\ \leggett,\ \Schmid,\ \Fisher,\ \Callan.  
The QFT \basa\ is integrable\ \GZ,\ \Warn\ and
many exact results, in particular concerning static transport properties, 
have been obtained or conjectured recently\ \Fen,\ \SalF,\ \BLZ.
In this paper we propose
another exact result for the model\ \basa, the expectation value of the 
exponential boundary field 
\eqn\resa{\eqalign{&\ \ \ \ \ \ \ \ \ \ \ \ \ \ \ \ \ \ \  
\langle\, e^{ia\varphi_{B}}\, \rangle=\bigg[\,
{2^{\beta^2} \pi\,\mu_{B}\over{\Gamma(\beta^2)}}\, \bigg]^
{{a^2}\over{1-\beta^2}}\times\cr&
\exp\biggl\lbrace\int_{0}^{\infty}{{dt}\over t}
\bigg[\ {{\big(\, e^{t}-1+e^{t (1-\beta^2)}+e^{-\beta^2 t}\, \big)\, 
\sinh^2 ( a\beta t )}\over{2\sinh(\beta^2 t)\, \sinh(t)\, 
\sinh\big((1-\beta^2)t\big)}}-a^2
\Big({1\over \sinh\big((1-\beta^2)t\big)}+
e^{-t}\Big)\ \bigg]\, \biggl\rbrace ,}}
with an arbitrary $a$ such that $|\Re e\,  2a| < \beta^{-1}$. 
In writing\ \resa\ we have assumed that the boundary
field $e^{ia\varphi_{B}}(x)$ is normalized in accordance with the short
distance limiting form of the two-point function
\eqn\dmnf{
\langle\,  e^{ia\varphi_{B}}(x)\,e^{-ia\varphi_{B}}(x')\, \rangle \to
|x-x'|^{-2a^2} \qquad {\rm as} \qquad |x-x'| \to 0\ ,}
so that the field $e^{ia\varphi_{B}}(x)$ has the dimension
$[\, mass\, ]^{a^2}$. The result\ \resa\ is
expected to hold in the domain
\eqn\djdhfg{\beta^2 < 1\ ,}
where the discrete symmetry $\varphi \to \varphi + 2\pi
n\beta^{-1}\ \ ( n=0,\pm1, \pm2...)$ of\ \basa\ is spontaneously
broken (equivalently, the ground state 
of the
associated quantum Brownian particle is localized) and by $\langle
... \rangle$ in\ \resa\ we mean the
expectation value over one of the ground
states, in which the field $\varphi_B (x)$ is localized near 0.

The model\ \basa\ has much in common with the
ordinary ``bulk'' sine-Gordon
model
\eqn\djfug{
{\cal A}_{SG} = \int_{-\infty}^{\infty}dx\,\int_{-\infty}^{\infty}dy\,
\bigg\{ {1\over {16\pi}}\Big[\,
\big(\partial_{y}\phi\big)^2 + \big(\partial_{x}\phi\big)^2\, \Big]
-2\mu\,\cos\big(\beta\phi\big)\bigg\}\ ,}
which can be thought of as the $c=1$ conformal field theory perturbed by
the ``bulk'' field $2\cos\big(\beta\phi\big)$.
The exact expectation
value of the ``bulk'' operator $e^{ia\phi}$ for the 
model\ \djfug\ (again,
in the domain\ \djdhfg), very similar to \resa,
has been proposed recently in \LZ.
We would like to note here that the model\ \djfug\ is
related by substitution
$\beta = ib, \ \mu\to-\mu$ to the sinh-Gordon model
\eqn\hasy{
{\cal A}_{ShG} = \int_{-\infty}^{\infty}dx\,\int_{-\infty}^{\infty}dy\,
\bigg\{{1\over {16\pi}}\Big[\, 
\big(\partial_{y}\phi\big)^2 + \big(\partial_{x}\phi\big)^2\, \Big]
+2\mu\,\cosh\big(b\phi\big)\bigg\}\ . }
The expression for $\langle\,  e^{ia\phi}\,
\rangle_{SG}$ in\ \LZ\ admits (for 
fixed $a$) a power
series expansion in $\beta^2$ with finite radius of convergence.
Therefore it is natural to assume that the vacuum expectation 
values of exponential fields in the sinh-Gordon model can be obtained
from the expression in\ \LZ\ simply
by the above continuation $\beta \to
ib$. This way one obtains
\eqn\ksjhh{\eqalign{
\langle\,  e^{a\phi}\,  \rangle_{ShG}=&
\bigg[\,
{m\,\Gamma\big(
{1\over 2+2b^2}\big)\,
\Gamma\big(1+{b^2\over 2+2b^2}\big)\over 4\sqrt{\pi}}
\, \bigg]^{-2a^2}
\times\cr
&\exp\biggl\lbrace\int_{0}^{\infty}{{dt}\over t}
\bigg[\ - {{\sinh^2 ( 2ab t )}\over{2\sinh(b^2 t)\, \sinh(t)\, 
\cosh\big((1+b^2)t\big)}}+
2a^2\,e^{-2t}\ \bigg]\, \biggl\rbrace\ ,}}
where\ \Zarn 
\eqn\sjusy{
m={4\sqrt{\pi}\over \Gamma\big({1\over 2+2b^2}\big)\,
\Gamma\big(1+{b^2\over 2+2b^2}\big) }\ 
\bigg[ -{\mu \pi \Gamma\big(1+b^2\big)\over \Gamma\big(-b^2\big)}\, 
\bigg]^{{1\over 2+2 b^2}}   }
is the particle mass of the  sinh-Gordon model. 

Unlike the sine-Gordon model\ \djfug\ it
is not very natural to think of\ \hasy\ as perturbed free boson conformal
field theory. Instead, the model\ \hasy\ is better understood in terms of 
the Liouville conformal field theory
\eqn\ssmnb{
{\cal A}_{Liouv}=
\int_{-\infty}^{\infty}dx\,\int_{-\infty}^{\infty}dy\,
\bigg\{{1\over {16\pi}}\Big[\, 
\big(\partial_{y}\phi\big)^2 + \big(\partial_{x}\phi\big)^2\,
\Big]
 + \mu\,e^{b\phi}\bigg\} }
perturbed by the operator $e^{-b\phi}$.
As is shown in\ \ZZ\ the operators
$e^{a\phi}$ in the Liouville theory\ \ssmnb\ satisfy
the ``reflection
relation''
\eqn\sbsf{e^{a\phi}(x,y) = R(a)\ e^{(Q-a)\phi}(x,y)\ ,}
with the function $R(a)$ related in a simple way to the Liouville
``reflection amplitude'' 
\eqn\skahg{R\big(Q/2 +iP\big) = S\big(P\big)=
-\bigg({{\pi\mu\, \Gamma(b^2)}\over{\Gamma(1-b^2)}}\bigg)^{-{2 iP\over b}}\
{{\Gamma\big(1+2iP/b\big)\, 
\Gamma\big(1+2iPb\big)}\over{\Gamma\big(1-2iP/b\big)\, 
\Gamma\big(1-2iPb\big)}} }
(see\ \ZZ\ for details). Here and below
\eqn\fkdja{Q=b^{-1}+b\ .}
It is not difficult to check that the one-point correlation
function\ \ksjhh\ in the sinh-Gordon model\ \hasy\ satisfies remarkable 
relation
\eqn\zmxbs{\langle\,  e^{a \phi}\,  \rangle_{ShG} = 
R(a)\ \langle\,  e^{(Q-a)\phi}\, \rangle_{ShG} }
with the same function $R(a)$ as in\ \skahg. Although at the moment we
do not have
completely satisfactory explanation for this phenomenon
\foot{Although the relation\ \zmxbs\ formally holds in the  conformal
perturbation theory for\ \hasy\ understood as
the Liouville theory perturbed
by $e^{-b\phi}$, this conformal perturbation theory
by itself does not give a 
valid definition of the one-point function\ \ksjhh.}, \zmxbs\ seems to
be a manifestation of an important hidden structure in the
sinh-Gordon theory
\foot{For instance, form-factors of the field $e^{a\phi}$ in the
sinh-Gordon model proposed in\ \KouM\ suggest that  matrix elements
$\langle\,  0\,  |\,  e^{a\phi}\,  |\, \theta_1 , ... , \theta_N\, 
\rangle $ (where
$|\, \theta_1 , ... , \theta_N\,  \rangle $ are  multiparticle states) 
satisfy the relation \zmxbs. Then it formally follows that
multipoint correlation functions, say $\langle\,  e^{a\phi}(x,y)\,
e^{a'\phi}(0,0)\, \rangle_{ShG} $, satisfy the relations similar
to\ \zmxbs, although this point needs further investigation. } and in
fact in more general Affine Toda field theories \foot{The relations
similar to\ \zmxbs\ are valid for  expectation values of exponential
fields in the Affine Toda theories and they can be used to determine
these expectation values\ \FATEEV.}. 
Let us note
that the relation\ \zmxbs\ together with an obvious symmetry
\eqn\erwe{\langle\,  e^{a\phi}\,  \rangle_{ShG} =
\langle\,  e^{-a\phi}\,  \rangle_{ShG} }
determines the expectation value $\langle\,
e^{a\phi}\, \rangle_{ShG}$ up to
a periodic function so that\ \ksjhh\ is a ``minimal solution'' to the
functional equations\ \zmxbs\ and\ \erwe. 

Let us come back to the boundary theory\ \basa. Consider the boundary 
sinh-Gordon model (BShG) which is obtained from\ \basa\ by substitution
$\beta = ib, \ \mu_B\to-\mu_B$,
just the same way\ \hasy\ is obtained from \djfug.
Again, BShG
can be thought of as the ``boundary Liouville theory''  
\eqn\dgfrs{{\cal
A}_{BLiouv}=
{1\over 4 \pi}
\int_{-\infty}^{\infty}dx\int_{0}^{\infty}dy\
\Big\{
\big(\partial_{y}\varphi\big)^2 + \big(\partial_{x}\varphi\big)^2\Big\}
+\mu_{B}
\int_{-\infty}^{\infty}dx \, e^{b\varphi_B(x)}\ .}
perturbed by the operator $e^{-b\varphi_B}$. As in the ``bulk''
Liouville theory, the boundary operators in the conformal
field theory\ \dgfrs\ satisfy the ``reflection relation''
\eqn\dkfhgt{e^{a\varphi_{B}}(x) =
R_{B}(a)\ e^{(Q-a)\varphi_{B}}(x) }
with $R_{B}(a)$ related to the associated ``reflection amplitude''
\eqn\sjhsgt{R_{B}\big(Q/2+iP\big) = S_{B}\big(P\big)=
\bigg(-{{2\pi\,\mu_{B}}\over{\Gamma(-b^2)}}\bigg)^{-{{2iP}\over{b}}}\
\bigg({b\over 2}\bigg)^{{2iP} Q}\ 
{{{\cal G}(-2iP)\,
{\cal G}^2 (Q/2+iP)}\over{{\cal G}(\ 2iP\,)\, {\cal G}^2
(Q/2-iP)}} \ .}
The function ${\cal G}(z)$ here is given by the integral
\eqn\sngdf{{\cal G}(z)=
\exp\biggl\{\int_{0}^{\infty} {d t\over t}\, \bigg[
{{e^{-{Q \over
2}t}-e^{-zt}}\over{(1-e^{-bt})(1-e^{-{t\over b}})}}+
{{Q\,(Q/2-z)\,e^{-{Q\over 2}t}}
\over{1-e^{-Qt}}}+{1\over 2}\,(Q/2-z)^2 \, e^{-{t\over 2}}\bigg]
\, \biggr\} }
in the domain $\Re e \,z > 0$ and it can be analytically continued into
the whole complex plane of $z$ by using the functional relations
\eqn\sksjssh{\eqalign{&{\cal G}(z+b)=
{1\over \sqrt{ \pi}}\  2^{ b (z+{Q\over 2})-1} \ 
b^{{1\over 2}-  bz}\ \Gamma\big( bz \big)\ {\cal G}(z)\ ,\cr
&{\cal G}\big(z+b^{-1}\big)=
{1\over \sqrt{ \pi}}\ 2^{{1\over b}(z+{Q\over 2})-1}\ 
b^{{z\over b}-{1\over 2}}\  \Gamma\big( b^{-1} z \big)\  {\cal G}(z)\ ,}}
which\ \sngdf\ satisfies.
It is easy to show that ${\cal G}(z)$ thus
defined is an entire function of $z$  with zeroes at $z=-nb-mb^{-1} \ (
 n,m = 0, 1, 2, ...)$.
The derivation of\ \dkfhgt,  \sjhsgt\ will be published
elsewhere. Now, let us assume that the vacuum expectation value
$\langle\,  e^{a\varphi_{B}}\, 
\rangle_{BShG}$ in the boundary sinh-Gordon
model satisfies a  ``reflection relation'' similar to\ \zmxbs, i.e.
\eqn\shdfaa{ \langle\,  e^{a\varphi_{B}}\,  \rangle_{BShG}=R_{B}(a)\,
\langle\,  e^{(Q-a)\varphi_{B}}\,  \rangle_{BShG}\ .}
It is not difficult to obtain the ``minimal'' solution to this 
functional equation which takes into account the symmetry relation 
analogous to\ \erwe. Then this solution can be continued back to pure 
imaginary $b = -i\beta$ which correspond to the boundary sine-Gordon
model\ \basa. This is the way we arrived at\ \resa.

In what follows we give some evidence in support of\ \resa.
First, the expectation value of the boundary field
$e^{i\beta\varphi_{B}}$ in the model\ \basa\ can be extracted from its
specific free energy
\eqn\sjshg{f_{B} =- \lim_{L\to\infty}\,L^{-1}\ \log Z_{L}\ , \qquad 
Z_{L} = \int\big[{\cal D}\varphi\big]\,e^{-{\cal A}_{BSG}}\ , }
where $L$ is the size of the system\ \basa\ in $x$ direction, namely
\eqn\hghsgf{\langle\,  e^{i\beta\varphi_{B}}\,  \rangle=
-{1\over 2}\,{\partial\over{\partial\mu_{B}}}\ f_{B}(\mu_{B})\ .}
The quantity $f_{B}$ for the boundary sine-Gordon model is known
exactly\ \Fen, \BLZ\ and\ \hghsgf\ gives
\eqn\msnsvb{\langle\,  e^{i\beta\varphi_{B}}\,  \rangle=
{{\Gamma\big({1-2\beta^2\over{2-2\beta^2}}\big)
\,\Gamma\big({\beta^2\over{2-2\beta^2}}\big)}
\over{4\,\pi^{3\over 2}\,(1-\beta^2)\,\mu_{B}}}\
\bigg[\, {{2\,\pi\,\mu_{B}}
\over{\Gamma(\beta^2)}}\, \bigg]^{1\over{1-\beta^2}}\ .}
It is easy to check that\ \resa\ agrees with\ \msnsvb . 

Expanding the expectation value $\langle\,  e^{ia\varphi_{B}}
\, \rangle$ into a power series in $a^2$ one can obtain the
expectation values of the polynomial fields
$\big(\varphi_{B}\big)^{2n}$. These in
turn admit expansions in power series in $\beta^2$.
In  this way one obtains
from\ \resa 
\eqn\smsh{\eqalign{
&\sigma_2 \equiv \langle\,  \varphi_{B}^2\,  \rangle= 
-2\,\log m - 2\,\gamma -
{\beta^4\over 6}\,  \big(\pi^2- 7 \zeta(3)\big)+O\big(\beta^6\big)\ ,\cr
&\sigma_4 \equiv 
\langle\,  \varphi_{B}^4\,  \rangle - 3\, \langle\,  \varphi_{B}^2
\, \rangle^2 =
4\,\beta^2
\, \big( \pi^2+7 \zeta(3)\big)+ 6\,\beta^4
\, \big( \pi^2+6 \zeta(3)\big)+O\big(\beta^6\big)\ ,\cr
&\sigma_6 \equiv 
\langle\,  \varphi_{B}^6\,  \rangle - 15\, \langle\,  \varphi_{B}^2
\, \rangle\ \langle\,  \varphi_{B}^4\,  \rangle+ 
30\,\langle\,  \varphi_{B}^2\,  \rangle^3 = - 16\,\beta^4 \, 
\big( \pi^4+93 \zeta(5)\big)+
O\big(\beta^6\big)\, .}}
where  $\gamma = 0.577216...$ is Euler's 
constant, $\zeta(s)$\ is the Riemann zeta function
and we have
introduced an
auxiliary mass parameter $m$ related to $f_{B}$ as
$$
m=- 2 \sin\Big({\pi\beta^2\over 1-\beta^2}\Big) \ f_B\  .
$$
On the other hand the coefficients in these expansions can be calculated
independently, from the standard perturbation 
theory for the action\ \basa. 
In this way one finds \foot{Our definition of the composite fields
$\big(\varphi_{B}\big)^{2n}$ which corresponds to the normalization
\dmnf\ 
is similar to that in the bulk sine-Gordon theory\ \LZ. For instance
$\varphi_{B}^2 (x) = \lim_{\varepsilon \to
0}\big[\, \varphi_{B}(x)\varphi_{B}(x+\varepsilon) +
2\log\varepsilon\, \big]$.}
\eqn\ndbfas{\eqalign{
&\sigma_2 = \lim_{\varepsilon\to
0}\Big[\, E(m\varepsilon) + 2\log \varepsilon\,  \Big] +\cr
&\ \ \ \ \ \ \ \ \ \ \ \ \ \ \ \ \ \ \ 
\ \ \ \ \ \ \   {\beta^4 \over 6}\, \biggl\{\,   
\int_{-\infty}^{\infty}{{d\rho}\over{2\pi}}
\int_{-\infty}^{\infty}{{d\rho'}\over{2\pi}}\ E(\rho) 
E(\rho') E^{3}(\rho-\rho')
-2\pi^2\,  \biggr\}+ O(\beta^6)\  ,\cr
&\sigma_4 = 
\beta^2\, 
\int_{-\infty}^{\infty}{{d\rho}\over{2\pi}}\  E^{4}(\rho) +
{3\over 2}\,\beta^4\  \int_{-\infty}^{\infty}{{d\rho}\over{2\pi}}
\int_{-\infty}^{\infty}{{d\rho'}\over{2\pi}}\  E^{2}(\rho) E^{2}(\rho')
E^{2}(\rho-\rho') +O(\beta^6)\ ,\cr
&\sigma_6 =\beta^4\ \biggl\{\, 10
\, \int_{-\infty}^{\infty}{{d\rho}\over{2\pi}}
\int_{-\infty}^{\infty}{{d\rho'}\over{2\pi}}\  E^{3}(\rho) E^{3}(\rho')
E(\rho-\rho')-\int_{-\infty}^{\infty}{{d\rho}\over{2\pi}}\  E^{6}(\rho)
\, \biggr\}  + O(\beta^6)
\ ,}}
where
\eqn\sjhg{
E(\rho) = \int_{-\infty}^{\infty} 
d\nu\  {{e^{i\nu\rho}}\over {|\nu|+1}}\ .}
Evaluating the integrals
in\ \ndbfas\ one finds perfect agreement with\ \smsh.

Finally, the following remark is in order. In the limit $\beta^2 \to 0$
the expectation value $\langle\,  e^{{\omega\over\beta}\varphi_{B}}\, 
\rangle$ in the  boundary 
sine-Gordon theory\ \basa\ can be described in terms of a particular
solution to the classical equations of motion corresponding to the
action\ \basa
\eqn\sjshy{
\big(\partial_{\sigma}^2 + \partial_{\tau}^2 \big)\Phi(\sigma,\tau) = 0 
\quad {\rm
for} \quad \tau > 0\  ; \qquad
\partial_{\tau}\Phi(\sigma,\tau)\big|_{\tau=0} =
\sin\Phi(\sigma,0)\ , }
where $\Phi(m_0 x, m_0 y)=\beta\varphi(x,y)$ and $
m_0 = 4\pi \beta^2\mu_{B}$. 
Let $\Phi(\sigma,\tau)$ be a 
function which solves
the equations\ \sjshy\ for $\tau\geq 0,\  \sigma^2+\tau^2 > 0$ and 
satisfies the 
following asymptotic conditions
\eqn\shsgf{\eqalign{&
\Phi(\sigma,\tau) \to 0 \quad {\rm as} \quad \sigma^2 + \tau^2 \to 
\infty\ ,\cr
&\Phi(\sigma,\tau) \to -\omega\log\big(\sigma^2+\tau^2\big) + 
C(\omega)\quad {\rm
as}\quad \sigma^2+\tau^2\to 0\ .}}
Here $C(\omega)$ is certain constant which in fact is completely
determined by the condition that such solution exists. It is easy to
show that the boundary value $\Phi_{B}(\sigma) = \Phi(\sigma,0)$ of 
this function
satisfies the integral equation
\eqn\xcdsh{\Phi_{B}(\sigma) =
\omega\,E(\sigma)+\int_{-\infty}^{\infty}{{d\sigma'}\over{2\pi}}\,
E(\sigma-\sigma')\big(\, \Phi_{B}(\sigma')-
\sin\Phi_{B}(\sigma')\, \big)\ ,}
where $E(\sigma)$ is the same function as in \sjhg. Therefore
\eqn\slskj{C(\omega) = -2\omega\gamma + 
\int_{-\infty}^{\infty}{{d\sigma}\over{2\pi}}\,
E(\sigma)\big(\, \Phi_{B}(\sigma)-\sin\Phi_{B}(\sigma)\, \big)\ .}
Now, the expectation value 
$\langle\,  e^{{\omega\over\beta}\varphi_{B}}\, \rangle$ with
fixed $\omega$
and $\beta^2 \to 0$ is expressed through the action\ \basa\ calculated 
on the above classical solution $\Phi(\sigma, \tau)$ which in
turn can be related to the constant $C(\omega)$
\eqn\slkjjhzn{\langle\,
e^{{\omega\over\beta}\varphi_{B}}\, \rangle \sim
m_0^{-{\omega^2\over\beta^2}}\  \exp\,\bigg(
\,{1\over{\beta^2}}\,\int_{0}^{\omega}{d\omega'}\,C(\omega')\,\bigg)\ .}
Our conjecture\ \resa\ allows us
to make the following prediction about the
constant $C(\omega)$
\eqn\slskjmu{
C(\omega) = -2\omega \gamma+\int_{0}^{\infty} {d t\over t}\ 
e^{t}\ \bigg[\, {2\omega t- \sin(2\omega t)\over\sinh^2(t)}\, \bigg]\ .}
It would be interesting to check this prediction by solving the
equation\ \xcdsh\ directly.

\centerline{}

\centerline{\bf Acknowledgments}
 
\hskip0.5cm
S.L. is grateful to the
Institute for Theoretical Physics at Santa Barbara for
their kind hospitality which he enjoyed during the
final stage of this work.
He also acknowledges stimulating  discussions with A. LeClair.
The work of S.L.
is supported in part by NSF under  grant No. PHY94-07194.
Research of A.Z.
is supported in part by DOE grant \#DE-FG05-90ER40559.
 
\listrefs

\end